\documentclass[preprint,12pt,3p]{elsarticle}

\usepackage{graphics}
\usepackage{amssymb}
\usepackage{amsthm}
\usepackage{amsmath}
\usepackage{float}
\usepackage{subcaption}
\usepackage{booktabs}
\usepackage{tabulary}
\usepackage{rotating}
\usepackage{setspace}
\usepackage{upgreek}
\usepackage{multirow}
\usepackage{pdflscape}
\usepackage[symbol]{footmisc}

\journal{Elsevier}

\begin{document}
	
\begin{frontmatter}
		\title{A New Class of Alumina-Forming Superalloy for 3D Printing}
		
		\author[1]{Joseph N. Ghoussoub}
		\author[1]{Przemys\l{}aw Klup\'{s}}
		\author[2]{William J B. Dick-Cleland} 
		\author[3]{Kathryn E. Rankin}
		\author[1]{Satoshi Utada}
		\author[1]{Paul A.J. Bagot}
		\author[1]{D. Graham McCartney} 
		\author[1]{Yuanbo T. Tang\correspondingauthor{}$^{,}$}
		\author[1,4]{Roger C. Reed}
		
		\address[1]{Department of Materials, University of Oxford, Parks Road, Oxford, OX1 3PH, UK}
		\address[2]{Alloyed Ltd. Yarnton, Kidlington OX5 1QU, UK}
		\address[3]{$\upmu$-VIS X-ray Imaging Centre, Faculty of Engineering and Physical Sciences, University of Southampton, SO17 1BJ, UK}
		\address[4]{Department of Engineering Science, University of Oxford, Parks Road, Oxford, OX1 3PJ, UK}
		
		\def\correspondingauthor{\footnote{Corresponding author: yuanbo.tang@materials.ox.ac.uk}}
		
\begin{abstract}
		A new class of crack-resistant nickel-based superalloy containing high $\gamma^\prime$ fraction is studied for the laser-powder bed fusion (L-PBF) process. The effects of the (Nb+Ta)/Al ratio is emphasised, a strategy that is shown to confer excellent low-temperature strength whilst maintaining oxidation resistance at high temperatures via stable alumina scale formation. The processability of the new alloys is characterised with respect to defect assessment by micro-focus  x-ray computed tomography; use is made of a prototype turbine blade geometry and the heritage alloy CM247LC as a benchmark. In all cases, some processing-related porosity is present in thin wall sections such as the trailing edge, but this can be avoided by judicious processing. The cracking seen in CM247LC -- in solid-state, liquation and solidification forms -- is avoided. A novel sub-solvus heat treatment strategy is proposed which takes advantage of AM not requiring solutioning; super-solvus heat treatment is inappropriate since it embrittles the material by deterioration of the texture and coarsening of grain boundary carbides. The tensile strength of the new superalloy is greatest when the Nb+Ta content is highest and exceeds that of CM247LC up to $\sim$900$\,$$^\circ$C. The oxidation resistance is best when Al content is highest, and oxidation-assisted cracking resistance maximized when the (Nb+Ta)/Al ratio is balanced. In all cases these are equivalent or superior to that of CM247LC. Nevertheless, the creep resistance of the new alloys is somewhat inferior to that of CM247LC for which the $\gamma^\prime$, C, and B contents are higher; this implies a processing/property trade-off which requires further clarification.
\end{abstract}
		
		
\end{frontmatter}
	
	\newpage
	
\section{Introduction}
	
One can argue that the Achilles' heel of 3D printing is the propensity for processing-related defects to occur	\cite{blakey2021metal}, due to the rapid cooling rates induced and thermal-cycling intrinsic to the process \cite{herzog2016additive}. This is particularly the case for the nickel-based superalloys studied here; modes of defect formation include porosity \cite{martin2019dynamics}, solidification cracking \cite{chauvet2018hot} and solid-state cracking \cite{qian2017statistical}. Such susceptibility is a disadvantage because conventional processing -- involving investment casting for example -- has evolved over many years to be a low and defect-free route with which 3D printing must compete \cite{pattnaik2012developments}. Its inherent advantages such as a strong coupling to computer-aided design technology help in this regard \cite{orme2017designing}.
	
What are the factors which might accelerate the insertion of 3D printing into a modern manufacturing economy? One approach involves tailoring the process through better understanding of the heat transfer and thermal-mechanical effects arising. The use of pre-heating \cite{ramirez2011novel} or multiple heat sources \cite{sanchez2021multi} are good examples of this first strategy. Alternatively, one can improve the input material employed as feedstock, arguing that a new process such as 3D printing will require new grades of alloy to be designed and proven \cite{pollock2020design}. There is evidence that this second approach has merit \cite{murray2020defect}. Alloy design approaches are emerging for this purpose \cite{zhou2020development, tang2021alloys, xu2020novel}. Nevertheless, work of the type reported here is needed to assess the effectiveness of the alloy design strategies. Are the compositions designed by the computer modelling truly superior? The only way to confirm this is via a vigorous and comprehensive programme of experimentation.
	
In this paper, we examine the processing and performance of a new grade of superalloy for high temperature applications. Three distinct attributes are considered: processability by laser-powder bed fusion (L-PBF), heat treatment response and finally mechanical behaviour. In this way, a holistic picture of the processing/property relationship is built up. We make use of a benchmark alloy, CM247LC, which is a widely available heritage alloy usually processed using conventional investment casting. The performances of three variants of the new alloy are compared with that of CM247LC. Consistent with the intended applications of these materials, the high temperature properties are emphasised particularly in creep and oxidation; the low temperature strength in the athermal regime is also considered.
	
	\newpage
	
\section{Experimental Methods}
	
\subsection{Alloy compositions and feedstock materials}
	
A new Ni-based superalloy \cite{nemeth2020patent} designed for additive manufacturing (AM) has been studied. The alloy was designed using the Alloys-by-Design (ABD) method \cite{reed2009alloys} allowing for a multi-objective optimization of various property trade-offs. Comprehensive descriptions of this alloy design methodology can be found in \cite{tang2021alloys, nemeth2020patent, reed2016alloys}. In this case, the alloy processability by L-PBF as well as the creep resistance, oxidation resistance, predicted maximal theoretical strength, and density have been considered. 

Three variants of the alloy named Alloy$\,$1, Alloy$\,$2, and Alloy$\,$3 were made available by Alloyed Ltd. for the present study on the basis of these criteria. These three compositions are calculated to have equal equilibrium $\gamma^\prime$ volume fraction at 900$\,$$^\circ$C, but possess varying (Nb+Ta)/Al ratios. This allowed for an experimental study of how the enrichment of $\gamma^\prime$ with either Al or (Nb+Ta) altered the microstructure-properties relationships. The equilibrium $\gamma^\prime$ volume fraction as a function of temperature for each composition was predicted by Thermo-Calc employing the TTNi8 database \cite{andersson2002thermo}.  Alloy$\,$3 had the greatest (Nb+Ta) content with reduced Al content, Alloy$\,$1 had greatest Al content with reduced (Nb+Ta), and Alloy$\,$2 had an intermediate (Nb+Ta)/Al ratio. These alloys were bench marked against the heritage CM247LC alloy.
	
The pre-alloyed argon gas atomised feedstock powders were prepared by Aubert \& Duval with median particle diameter (D50) between 32.3 and 33.0$\,$$\upmu$m. The D10 and D50 values were  $\sim$19 and 55$\,$$\upmu$m respectively. The measured alloy powder compositions, as determined by inductively coupled plasma-optical emission spectroscopy (ICP-OES) and ICP-combustion analysis, are given in Table~\ref{table1}. The powder particles were predominantly spherical, with a small number of satellites and other morphological irregularities present.
	
	\begin{landscape}
	\begin{table}[H]
		\centering
		\caption{Compositions of the alloy powders at\%.}
		\begin{tabular}{@{}cccccccccccccc@{}}
			\toprule
			Alloy & Ni   & Al   & Co   & Cr  & Mo   & Nb  & Ta   & Ti   & W   & C    & B    & Hf   & (Nb+Ta)/Al \\ \midrule
			Alloy$\,$1 & bal. & 9.5  & 33.4 & 8.4 & 0.62 & 1.1 & 0.89 & 0.12 & 1.8 & 0.15 & 0.02 & - & 0.21   \\
			Alloy$\,$2 & bal. & 8.6  & 33.5 & 8.5 & 0.62 & 1.5 & 1.24 & 0.13 & 1.9 & 0.15 & 0.02 & - & 0.32  \\
			Alloy$\,$3 & bal. & 7.8  & 34.2 & 8.4 & 0.64 & 1.9 & 1.53 & 0.13 & 1.9 & 0.13 & 0.02 & - & 0.43   \\
			CM247LC   & bal. & 10.9 & 17.8 & 8.5 & 0.31 & - & 0.95 & 0.82 & 2.8 & 0.31 & 0.10 & 0.42 &        \\ \bottomrule
		\end{tabular}
		\label{table1}
	\end{table}
\end{landscape}	
	
\subsection{Processing by additive manufacturing}
	
L-PBF was carried out by Alloyed Ltd. using a Renishaw AM 400 pulsed fibre laser system of wavelength 1075$\,$nm under an argon atmosphere with a build plate size of 80$\times$80$\times$64$\,$mm$^3$. The processing parameters employed were: laser power 200$\,$W,  laser focal spot diameter 70$\,$$\upmu$m, powder layer thickness 30$\,$$\upmu$m, and pulse exposure time 60$\,$$\upmu$s. A ‘meander’ laser scan path pattern was used with hatch spacing of 70$\,$$\upmu$m and laser scan speed of 0.875$\,$m/s, the path frame of reference was rotated by 67$\,$$^{\circ}$ with each layer added. In order to produce a high quality surface finish, the laser traced the border of the sample after each layer, the laser speed on the borders was reduced to 0.5$\,$m/s. Each alloy powder was processed with these parameters. These were selected on the basis of a previous study of the influence of processing conditions on cracking in CM247LC \cite{ghoussoub2020influence}.
	
For each alloy, the same build plate configuration was employed comprising several geometries. Cubes of dimensions 10$\times$10$\times$10$\,$mm$^3$ were printed for microstructural characterization and for the preparation of oxidation test coupons. Vertical bars of dimensions 10$\times$10$\times$52$\,$mm$^3$ were printed and were used for the machining of mechanical test pieces. These bars were manufactured with 16 inverted pyramid legs in order to allow for easy removal from the baseplate. Discs of diameter 3$\,$mm and height 1$\,$mm were produced for differential scanning calorimetry (DSC). Turbine blade shaped-parts were printed to simulate the manufacture of an engineering component; in this case representative of those used in a helicopter engine, these had a height of 30$\,$mm.
	
\subsection{Materials characterization}
	
\subsubsection{Optical and scanning electron microscopy (SEM)}
	
Optical microscopy was used to assess the cracking susceptibility. The cube samples were sectioned on planes parallel (termed XZ or YZ) and perpendicular (termed XY plane) to the build direction. Thus with this definition, the XY plane has as its normal the build direction. For the XY plane, a section was taken at the mid-height. The severity of cracking was determined by imaging the XY plane; 5 images were considered per alloy. The ImageJ software \cite{abramoff2004image} was used to compare the alloys by determining the crack count density (cracks/mm$^2$) and crack length density (mm/mm$^2$) -- note that the crack length was defined as the caliper diameter corresponding to the largest line length across each crack. The microstructure was investigated in the as-printed, heat treated, and oxidized condition using backscattered electron (BSE) imaging and energy dispersive X-ray spectroscopy (EDX) in a Zeiss Merlin Gemini 2 field emission gun scanning electron microscope (FEG-SEM) equipped with an Oxford Instruments XMax 150mm/mm$^2$ energy-dispersive detector and a  electron backscattered diffraction (EBSD) system.  EDX maps were acquired using accelerating voltage of 10$\,$kV and probe current 500$\,$pA. Samples were electrolytically etched to remove the $\gamma$ matrix using 10\% phosphoric acid at 3$\,$V direct current in order to observe the $\gamma^\prime$ size, distribution, and morphology.

\newpage
	
\subsubsection{Atom probe tomography (APT)}
	
The Alloy$\,$1 and Alloy$\,$3 cubes were cut into rectangular prism matchsticks of dimensions 10$\times$1$\times$1$\,$mm$^3$ using a Struers Accutom-50 and alumina saw. These were prepared for atom probe analysis through electrochemical polishing in a solution of 25$\,$\% perchloric acid and 90$\,$\% acetic acid with voltage of 25$\,$V. A second stage of polishing to finalize the tip was done in a solution of 2$\,$\% perchloric acid and 98$\,$\% butoxyethanol with voltage of 20$\,$V. Specimens were analysed using a Cameca LEAP 5000XR system with detection rate of 52$\,$\% and laser wavelength 355$\,$nm. Samples were cryogenically cooled to 50$\,$K inside an ultra high vacuum chamber at $<$ 4$\times$10$^{-11}$$\,$Torr. Laser mode was employed at a pulse frequency of 200$\,$Hz and energy 50$\,$pJ. The datasets were reconstructed using Integrated Visualization and Analysis Software (IVAS). Due to the laser mode, the crystallographic poles could not be used for more accurate spatial calibration. The parameters used were image compression factor (ICF) of 1.65, and an initial cap radius of 35 and 25$\,$nm as well as geometric field-factor of 3.3 and 4.5 for Alloy$\,$1 and Alloy$\,$3 respectively. The $\gamma$-$\gamma^\prime$ interface of primary $\gamma^\prime$ precipitates was analysed in tips produced from Alloys$\,$1 and 3 using cuboid regions of interest that have axes normal to the interface. These volumes were divided into bins of width of 0.5$\,$nm and the composition of each bin was calculated using AtomprobeLab software and the error of the composition of each element was calculated as established in \cite{london2019quantifying}. 
	
\subsubsection{X-ray computed tomography (XCT)}
	
X-ray computed tomography of the turbine blade-shaped samples was carried out with a custom Nikon XTEK XTH 225 kVp micro-focus CT system fitted with a 2000$\times$2000 pixel Perkin Elmer XRD 1621 CN14 HS Detector (PerkinElmer Optoelectronics, Germany). In order to detect the micron length scale defects characteristic of L-PBF a slice of the top 1$\,$mm of the turbine blade was machined by wire electrical discharge machining (EDM) for analysis. Three  separate XCT scans were performed on each sample which were concatenated together to maximise the voxel (volume element or cubic pixel) resolution achievable for the entire length. 

Scans were performed at 210$\,$kVp peak voltage and 42$\,$$\upmu$A current, with a source to object distance of 17$\,$mm and an source to detector distance of 797$\,$mm. Using an analogue gain of 24$\,$dB, 3142 projection images were acquired throughout 360 degrees rotation, averaging 8 frames per projection with 1$\,$s exposure time per frame.  Projection images were reconstructed into 32 bit float volumes using filtered back-projection algorithms implemented within CTPro3D and CTAgent software v2.2 (Nikon Metrology, UK). These were downsampled to 8 bit (to reduce processing time), manually thresholded and converted to binary. The defects were quantified in terms of size distribution. The size was defined by defect volume using the 3D objects counter in ImageJ. A further analysis of a $\sim$500$\times$500$\times$500$\,$$\upmu$m$^3$ sub-volume in the bulk of the CM247LC was carried out to understand the crack morphology in 3D.
	
\newpage	
	
\subsection{Heat treatment}
	
In order to design suitable heat treatments, differential scanning calorimetry (DSC) was first used to determine the heat treatment window. A NETZSCH 404 F1 Pegasus instrument with a nitrogen cover gas flowing at 50$\,$ml/min was employed. Samples were heated to 700$\,$$^\circ$C at 20$\,$K/min, then from 700$\,$$^\circ$C to 1450$\,$$^\circ$C at 10$\,$K/min. The $\gamma^\prime$ solvus temperature was determined by the inflexions in the DSC signal following Chapman and Quested \cite{chapman2004application, quested2009measurement}. The $\gamma^\prime$ solvus temperatures were 1172, 1179, 1182, and 1249$\,$$^\circ$C for Alloy$\,$1, Alloy$\,$2, Alloy$\,$3, and CM247LC respectively.
	
Super-solvus and sub-solvus heat treatments (HT) designated HT$\,$1 and HT$\,$2 respectively were performed on Alloy$\,$2 on the basis of the measured $\gamma^\prime$ solvus temperature. A further sub-solvus single step heat treatment designated HT$\,$3 was performed to assess the feasibility of achieving properties with reduced time. These three heat treatments are summarized in Table~\ref{table2}, each heat treatment step was followed by air cooling (AC). 
	
\begin{table}[H]
	\centering
	\caption{Heat treatments applied to Alloy$\,$2.}
	\begin{tabular}{@{}cccc@{}}
		\toprule
		Process             & Stage 1      & Stage 2      & Stage 3       \\ \midrule
		HT 1 (super-solvus) & 1220$\,$$^\circ$C 2$\,$h AC & 1100$\,$$^\circ$C 4$\,$h AC & 850$\,$$^\circ$C 20$\,$h AC \\
		HT 2 (sub-solvus)   & 1100$\,$$^\circ$C 4$\,$h AC & 850$\,$$^\circ$C 20$\,$h AC             & -             \\
		HT 3 (sub-solvus)   & 1080$\,$$^\circ$C 4$\,$h AC & -            & -             \\ \bottomrule
	\end{tabular}
	\label{table2}
\end{table}	
	
\subsection{Oxidation study}
	
Thermo-gravimetric analysis (TGA) was used to assess the oxidation resistance of the alloy, using a NETZSCH STA 449 F1 Jupiter. Specimens of dimensions 10$\,$mm$\times$10$\,$mm$\times$1$\,$mm were cut perpendicular to the build direction  and polished to a mirror finish with 4000 grit SiC paper. The tests were carried out in 5 steps, initially a protective Ar gas was run for 1$\,$h to ensure equilibrium of the system. This was followed by an increase in temperature at a rate of 20$\,$K/min to the test temperature of 1000$\,$$^\circ$C which was then followed by a hold for 30$\,$min under protective Ar and then the onset of the oxidizing laboratory air flow at a rate of 50$\,$ml/min for 24$\,$h. After this period the sample is cooled to room temperature at 20$\,$K/min. The evolution of mass gain was analysed starting at the onset of oxidizing air. A set of 10 BSE images were taken of the oxide layer and $\gamma^\prime$ depletion zone in each alloy to quantify their thickness and size using a macro in ImageJ. 
	
\newpage	
	
\subsection{Thermo-mechanical testing}
	
\subsubsection{Tensile properties}
	
Isothermal uniaxial tensile tests were performed using a 5$\,$kN Instron electro-thermal mechanical testing (ETMT) machine. Specimens of full length 40$\,$mm, gauge length 14$\,$mm, and 1$\,$mm$^2$ cross sectional area were machined with axis along the build direction from the additively manufactured bars by wire electrical discharge machining (EDM), and polished to mirror finish 4000 grit SiC paper to negate the influence of EDM induced surface defects. The strain was measured via a non-contact iMetrum digital image correlation system. Joule heating under free expansion conditions was employed to reach the test temperatures using a heating rate of 200$\,$K/s. The temperature was measured by spot-welding a K-type thermocouple to the center of specimen gauge.
	
The 4 compositions were tested at room temperature and in the range of 600-1100$\,$$^\circ$C in increments of 100$\,$$^\circ$C. Specimens were strained rapidly at a rate of $10^{-2}$$\,$s$^{-1}$ to reduce the influence of dynamic precipitation during testing and facilitate high-throughput experimental work. At all temperatures the flow stress respectively were taken as the 0.2$\%$ proof stress. The sub-solvus and super-solvus properties of Alloy$\,$2 were additionally compared across the temperature range by testing at a strain rate of $10^{-3}$$\,$s$^{-1}$.
	
\subsubsection{Creep testing}
	
For each alloy, bars of dimensions 10$\times$10$\times$52$\,$mm$^3$ were heat treated and machined into specimens for creep testing. Bars were heat treated at 1080$\,$$^\circ$C for 4$\,$h and air cooled. The selection of this sub-solvus heat treatment is elucidated in the results section. Bars were machined and tested externally by the accredited laboratory Westmoreland Mechanical Testing \& Research, Ltd in accordance to ASTM E139. Samples were machined with gauge length of 20$\,$mm and full length of 52$\,$mm. Creep tests were performed at temperatures ranging between 800$\,$$^\circ$C and 1050$\,$$^\circ$C at various stress levels between 180$\,$MPa and 500$\,$MPa
	
\subsubsection{Resistance to oxidation-assisted cracking (OAC)}
	
In order to assess the capability of the alloys to withstand oxidation-assisted cracking (OAC), specimens were tensile tested by ETMT at strain rates of $10^{-2}$$\,$s$^{-1}$, $10^{-3}$$\,$s$^{-1}$, and $10^{-5}$$\,$s$^{-1}$ in laboratory air at 800$\,$$^\circ$C. This methodology, developed by N\'{e}meth et al \cite{nemeth2017environmentally}, allows for the assessment of the influence of oxidation on crack growth which is critical for high temperature component design \cite{li2015effects}.
	
\newpage	
	
\section{Results}
	
\subsection{Resistance to processing induced defects}
	
With regard to processability, the situation has been found to be very complicated so in this first introductory paragraph -- in order to improve readability -- an overview is first provided; further confirmatory evidence is provided in the following two paragraphs. 

\begin{landscape}
	\begin{figure}[H]
		\centering
		\includegraphics[width=0.95\columnwidth]{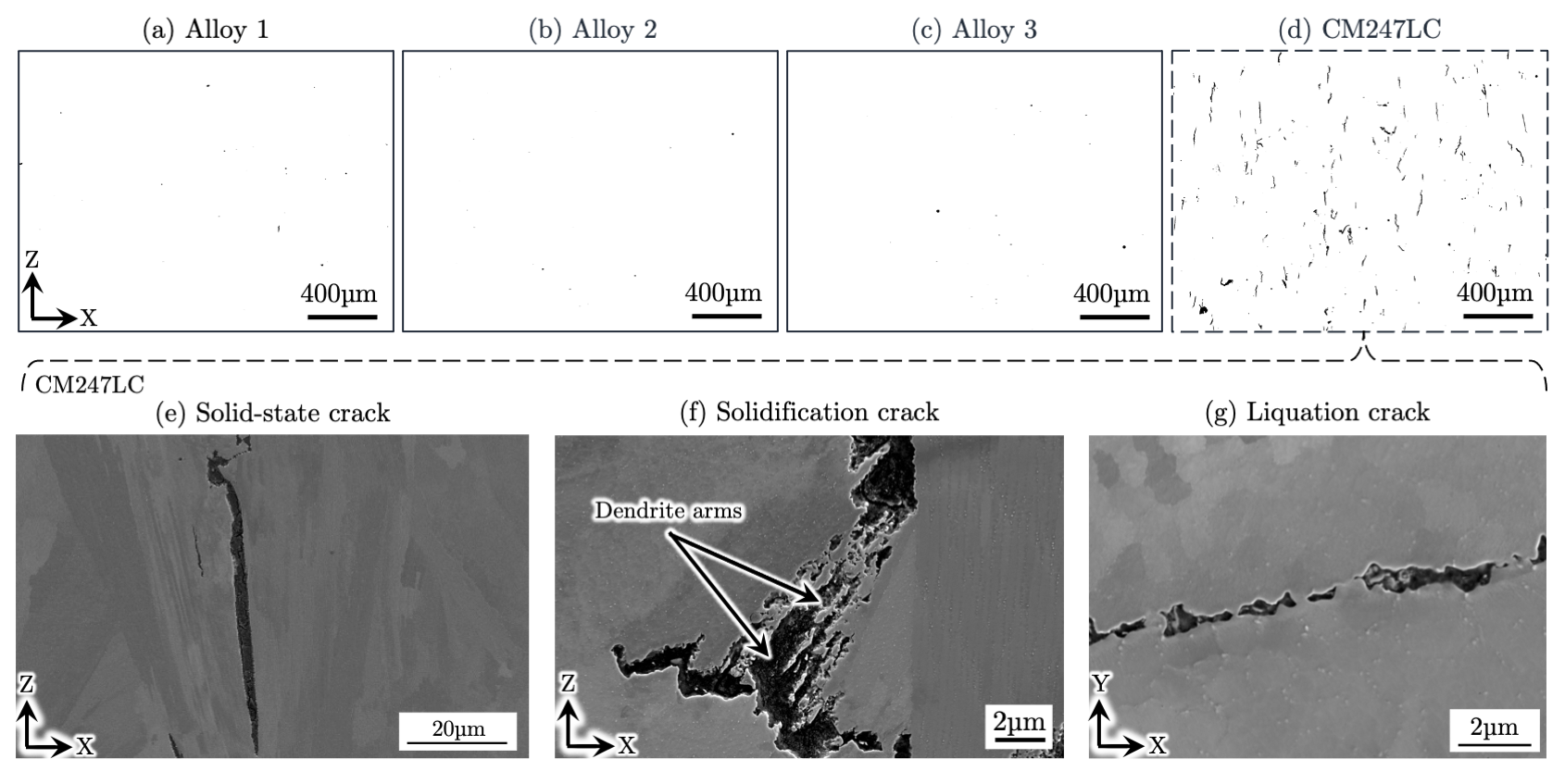}
		\caption{(a-d) Optical micrographs with black and white threshold applied showing the cracking in CM247LC vs lack thereof in Alloy variants$\,$1-3. SEM micrographs showing examples of (e) solid-state cracking (f) solidification and (g) liquation cracking observed in CM247LC.}
		\label{figure1}
	\end{figure}
\end{landscape}

Consider Figure~\ref{figure1} which contains optical micrographs taken within the bulk of XZ plane cross-sections -- containing the build direction -- obtained from printed cubes. These micrographs illustrate that the new compositions print with no evidence of defects, at least within the bulk, cracks were not detected across an area of 100$\,$mm$^2$ at a magnification of 100$\times$. On the other hand, alloy CM247LC was found to contain crack-like defects due to processing: measured were a crack count density of $\sim$49$\,$cracks/mm$^2$ and crack length density of 2.6$\,$mm/mm$^2$. This disparity in behaviour between the new and CM247LC is further evidence for the strong composition dependence of cracking in L-PBF processing consistent with observations reported elsewhere \cite{ghoussoub2021influence}. Nevertheless, for all compositions examined here -- both CM247LC and the novel compositions -- the edges of the built structures contained gas-related porosity which had survived the harsh thermal-fluid flow induced by the laser/material interaction and indeed the higher energy density processing which was carried out at these locations. This suggests that the formation of such gas porosity is not strongly related to alloy composition but rather to processing conditions. In what follows below, further evidence is provided for purposes of corroboration. 

\begin{figure}[H]
	\centering
	\includegraphics[width=0.88\columnwidth]{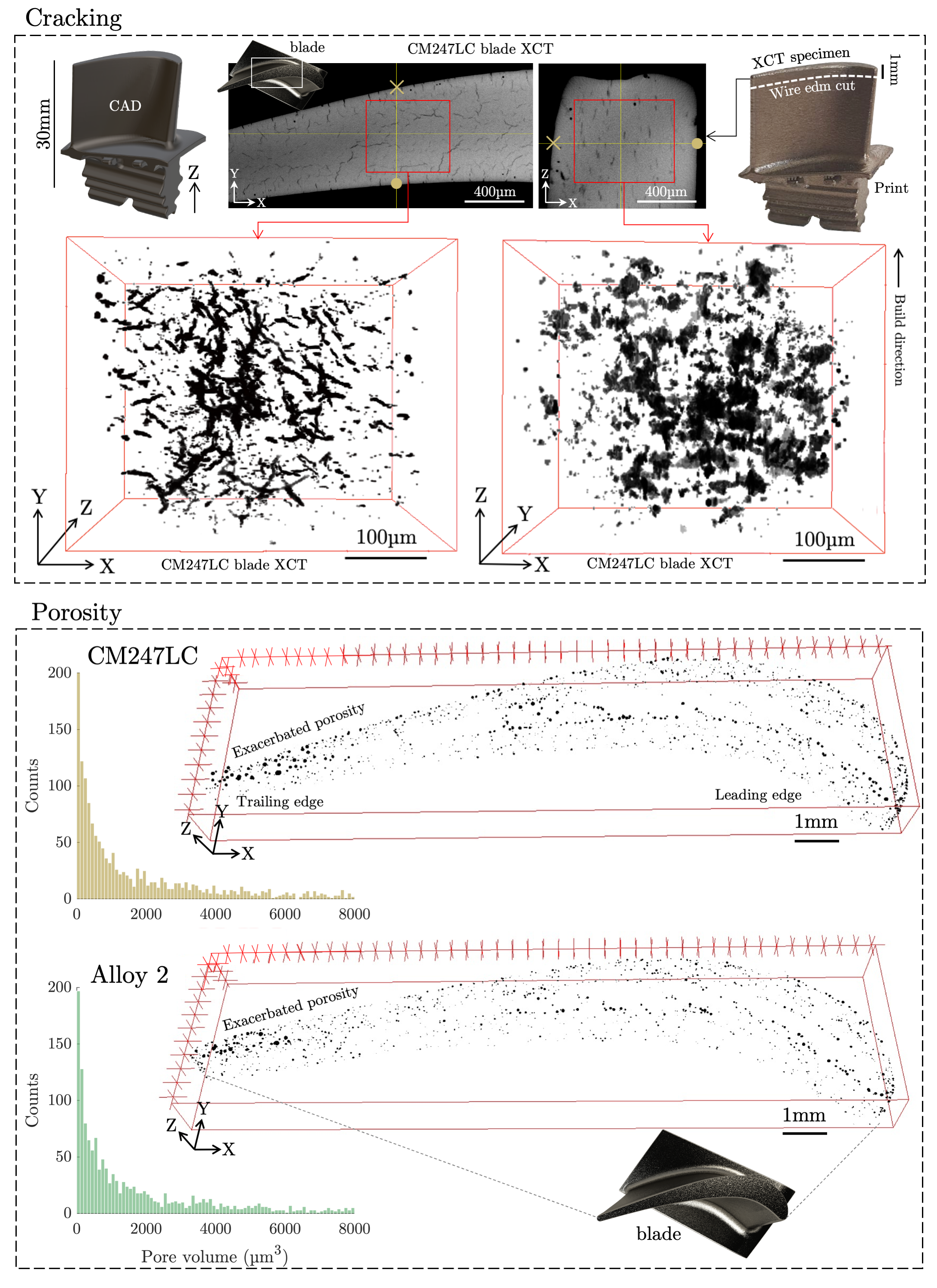}
	\caption{ Slices of the CM247LC blade sample XY and XZ planes as observed by XCT and 3D reconstruction of cracks as observed on each plane. Quantification of porosity and 3D reconstruction showing where porosity occurred in CM247LC and Alloy$\,$2.}
	\label{figure2}
\end{figure}

With regard to the crack-like defects in CM247LC, optical microscopy on the 2D sections XZ confirm a strong directionality, with the cracks appearing largely straight and aligned in the Z build direction. Further characterisation by SEM indicates that cracks formed by three mechanisms: solid-state, solidification, and liquation  cracking. Examples of these types of cracks is shown in Figure~\ref{figure1}. The presence of solid-state cracks is strongly evidenced by their length being greater than the dimensions of a single melt pool -- the use of the pulsed laser system employed here results in discretized melting events. These melt pools were determined to have radii of $>$70$\,$$\upmu$m; on the basis of this it was concluded that cracks of length greater than 70$\,$$\upmu$m propagated at least in part in the solid-state. The occurrence of solidification cracks is supported by the presence of retained dendritic features, which confirm that separation occurred in the presence of liquid phase. Lastly, the manifestation of liquation cracking is suggested by $\upmu$m length scale cracks with smoothed edges indicative of a solidified liquid film \cite{tang2021alloys}. Characterisation using micro-XCT on the printed turbine blades has helped to elucidate the situation further, by revealing the 3D nature of cracking in CM247LC. Cracks were not detected by XCT for any grades of the new alloy, but for CM247LC cracking was detected by XCT throughout the entirety of top 1mm of the blade profile characterised, as shown in Figure~\ref{figure2}. Analysis of the sub-volume in the CM247LC blade revealed the 3D shape of the cracks; when observed on the XY plane they appear skeleton-like, whereas when observed on the XZ or YZ plane they appear plate-like. This supports observations made by optical microscopy and arises because cracking occurs along high angle grain boundaries (HAGB) \cite{chauvet2018hot, hariharan2019misorientation} and the microstructure produced by AM is highly textured with grains elongated in the  build direction (Z-axis).

We now turn to the gas-related porosity. Quantification of the XCT data shows porosity occurred to the same extent and in both new alloys and CM247LC. The total number of pores, mean, median, and max pores volumes were 1561, 3690$\,$$\upmu$m$^3$, 1090$\,$$\upmu$m$^3$, and 15590$\,$$\upmu$m$^3$ for Alloy$\,$2 and 1860, 4350$\,$$\upmu$m$^3$, 885$\,$$\upmu$m$^3$, and 18920$\,$$\upmu$m$^3$ for CM247LC. Porosity occurred largely at the sample edges, hence it provides a delineation of the outline of the blade when visualized in 3D. In all cases here, porosity was exacerbated at the trailing edge and leading edge. The influence of processing conditions on porosity has been widely reported \cite{martin2019dynamics, khairallah2016laser, moussaoui2018effects} where increased energy density in particular can cause keyholing and entrainment of the chamber cover gas \cite{panwisawas2017keyhole}. The reduced scan velocity used at the sample edges -- selected to improve the sample surface finish -- is possibly a contribution to the porosity observed. Furthermore, the thin sections at the trailing edge have incurred even greater gas porosity due to the beam turning corners frequently and increasing further the mean energy density \cite{catchpole2017fractal, comminal2019motion}. The spatial dependence of defects in the blades confirms geometric changes -- even in a small blade -- have profound influence on the local formation of defects. Hence, the influence of changing geometry suggests that forgiving compositions that are processable across a wide range of heat transfer conditions are required to account for geometric effects in actual engineering components.

\subsection{On the influence of heat treatment and microstructure}
	
The Ni-based superalloy microstructure following printing consisted of cellular dendrites with small or no secondary dendrite arms. The intercellular enrichment of $\gamma^\prime$ formers and intercellular precipitation of Ta and Nb-rich MC carbides was observed; in CM247LC, Hf-rich carbides also precipitated at the cell boundaries, consistent with Ref. \cite{divya2016microstructure, wang2017microstructure}. Despite their substantial equilibrium $\gamma^\prime$ volume fractions, the novel alloy variants and CM247LC printed to $\gamma$, as observed on the lengthscale detectable by SEM. While it is possible that nm length scale $\gamma^\prime$ -- not observable by SEM -- precipitated by reheating from subsequent passes, these findings emphasize that heat treatment is needed in order to develop desirable mechanical properties.

\begin{figure}[H]
	\centering
	\includegraphics[width=1\columnwidth]{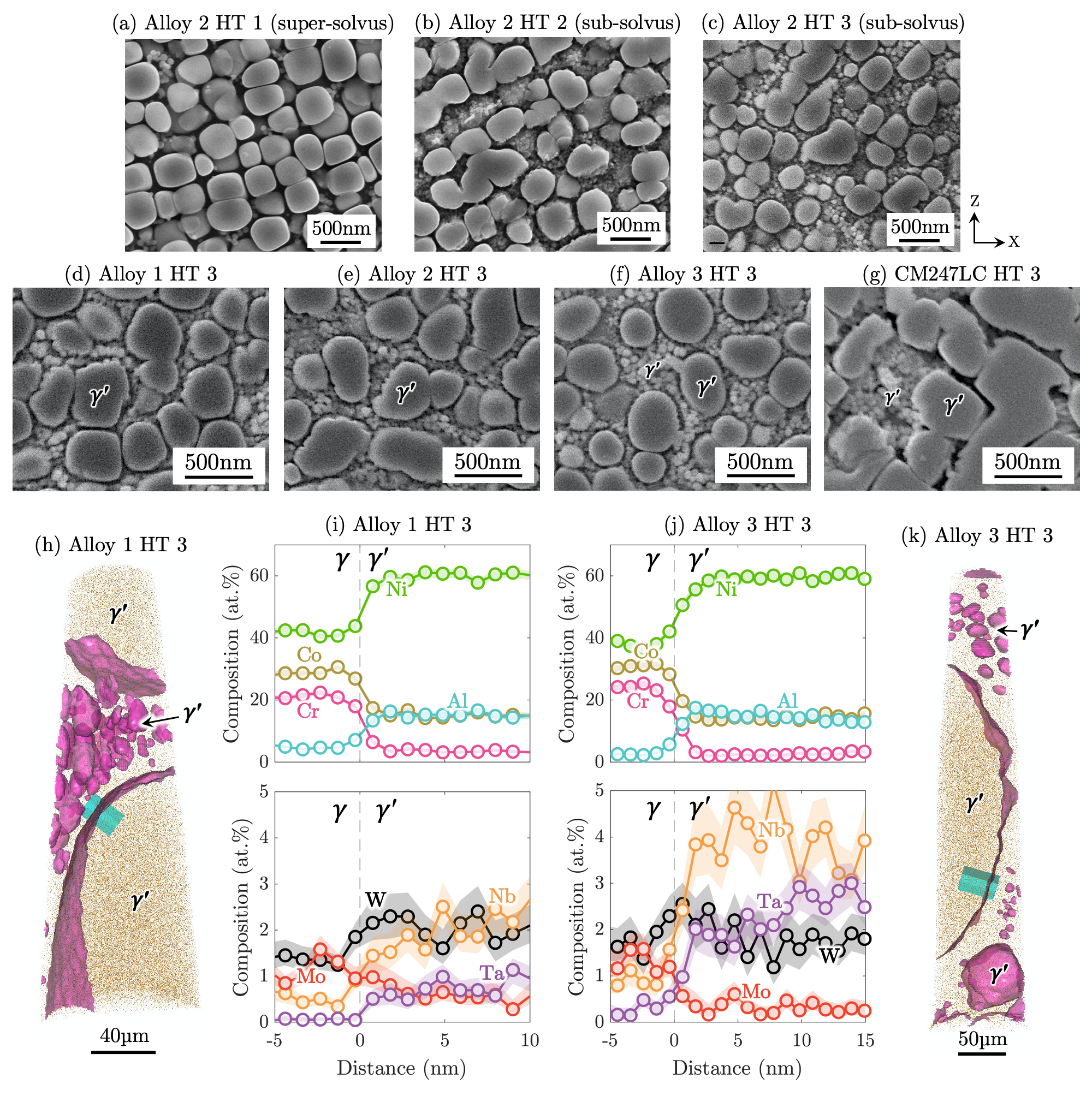}
	\caption{SE micrographs showing the $\gamma^\prime$ inside the grain of the Alloy$\,$2 microstructure following (a) HT$\,$1 (b) HT$\,$2 (c) HT$\,$3 and of (d) Alloy$\,$1 (e) Alloy$\,$2 (f) Alloy$\,$3 and (g) CM247LC after HT$\,$3. Migrographs were all taken on the XZ plane. The 3D reconstruction of the atom probe tip of (h) Alloy$\,$1 and (k) Alloy$\,$3 following HT$\,$3 showing 13.5$\,$at\% Cr iso-surfaces around $\gamma^\prime$ precipitates as well as the cuboid regions of interest used to extract concentration profiles across the $\gamma$-$\gamma^\prime$ interface. The concentration profiles across the $\gamma$-$\gamma^\prime$ interface are shown for (i) Alloy$\,$1 and (j) Alloy$\,$3.}
	\label{figure3}
\end{figure}

The SEM micrographs in Figure~\ref{figure3} show the microstructure developed in Alloy$\,$2 following HT$\,$1-3 and the microstructure of the 4 compositions considered in this work following HT$\,$3. The $\gamma^\prime$ distribution of Alloy$\,$2 was unimodal following HT$\,$1 and bimodal following HT$\,$3. The primary $\gamma^\prime$ observed in CM247LC following HT$\,$3 was larger than in the new compositions. 

Primary and secondary $\gamma^\prime$ was observed in all the alloys and APT analysis of Alloys$\,$1 and 3 following HT$\,$3 demonstrated the presence of tertiary $\gamma^\prime$. APT confirmed the segregation of Ta, Nb, Al, and Ni to $\gamma^\prime$. Conversely, Co, Cr, and Mo segregated to $\gamma$. No clear partitioning was observed for W. Crucially, the concentrations of Nb and Ta in the $\gamma^\prime$ of Alloy$\,$3 were approximately 4$\,$at.\% and 2.5$\,$at.\% respectively, these were twice as high as those of Alloy$\,$1. The macroscopic influence of this microscopic change is elucidated further sections.

\begin{figure}[H]
	\centering
	\includegraphics[width=1\columnwidth]{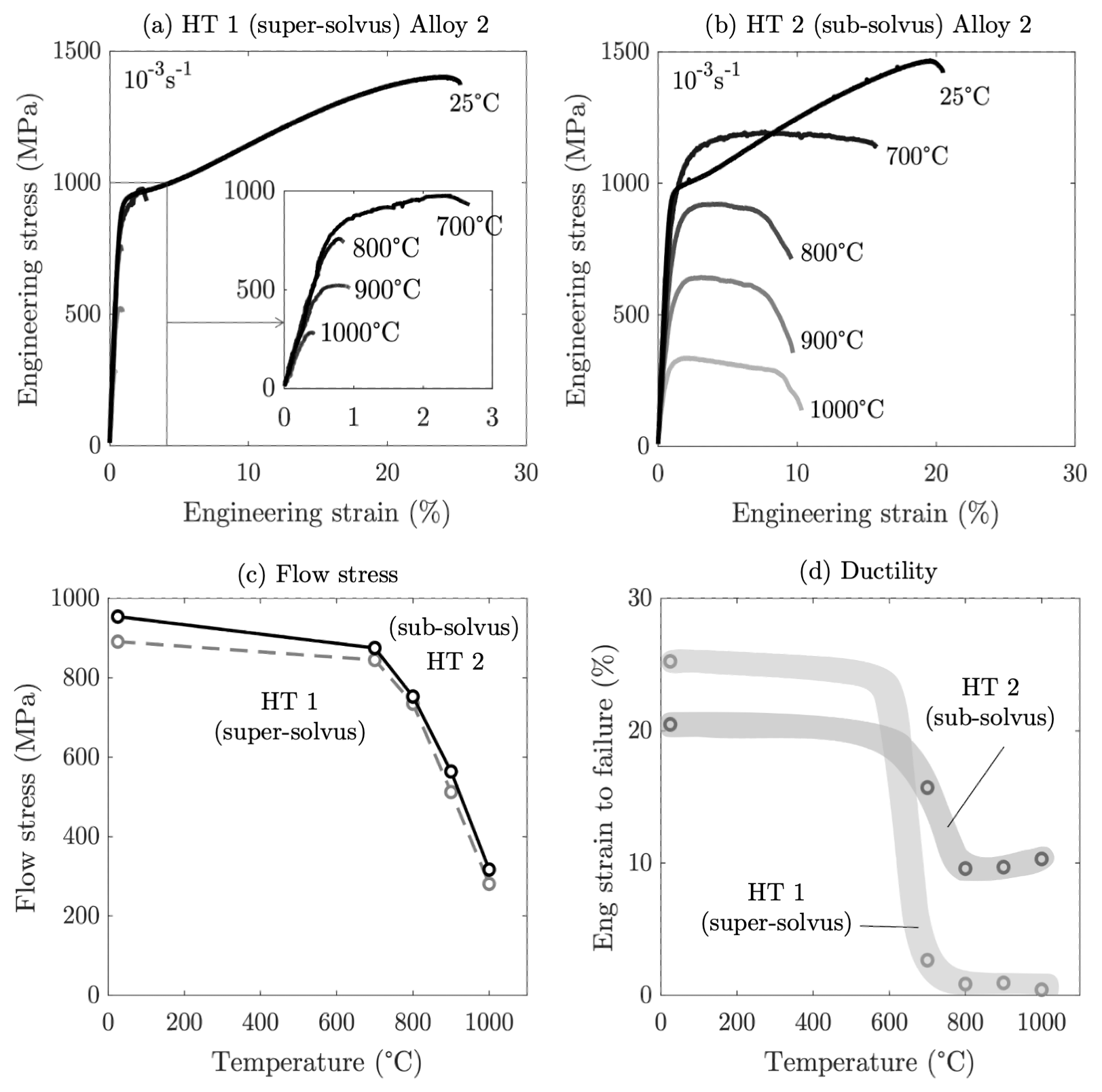}
	\caption{Tensile response of Alloy$\,$2 at various temperatures following (a) HT$\,$1 (super-solvus) and (b) HT$\,$2 (sub-solvus), as well as their summarized (c) flow stress and (d) ductility across the temperature range.}
	\label{figure4}
\end{figure}

When tensile tested at a strain rate of 10$^{-3}$$\,$s$^{-1}$ Alloy$\,$2 exhibited greater high temperature ductility and similar flow stress following the sub-solvus HT$\,$2 -- as compared to the super-solvus HT$\,$1. When Alloy$\,$2 was heat treated in the super-solvus condition the material exhibited brittleness, accommodating only $\sim$1$\,$\% ductility when tested between 700$\,$$^\circ$C-1000$\,$$^\circ$C, see Figure~\ref{figure4}. The root cause of the embrittlement following HT$\,$1 is discussed in Section~4. But here, it is sufficient to note for now that the flow stress of Alloy$\,$2 at 800$\,$$^\circ$C was 30$\,$MPa greater when processed by HT$\,$3 in contrast to HT$\,$2. In light of this, as well as the 20$\,$h reduction in processing time, HT$\,$3 was used to process the alloys for comparison of their properties in all work that follows.

\newpage	
\subsection{On the alloy properties developed following heat treatment$\,$3 (HT$\,$3)}
	
\subsubsection{Flow behaviour}
	
\begin{figure}[H]
	\centering
	\includegraphics[width=0.95\columnwidth]{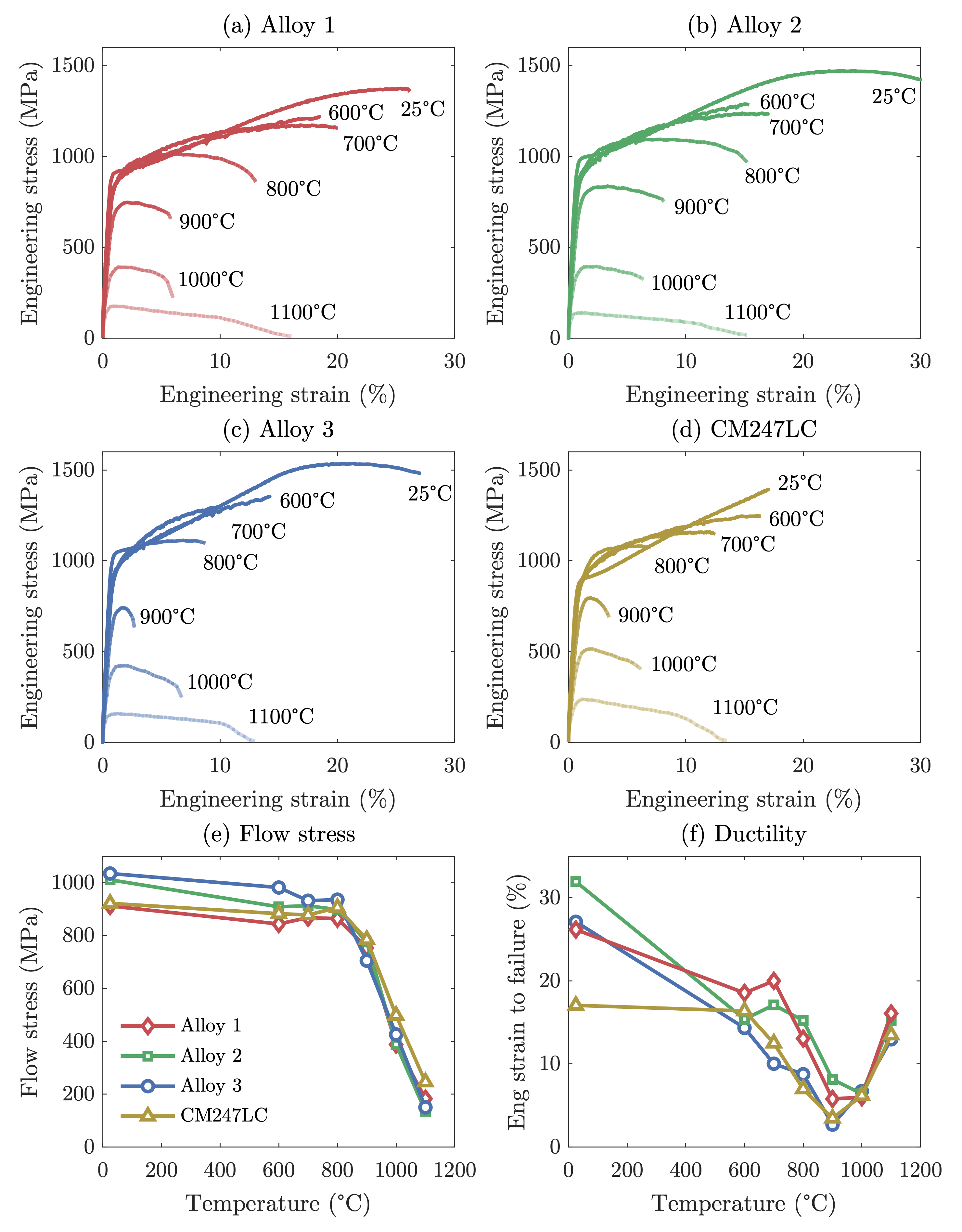}
	\caption{Tensile response after HT$\,$3 for (a) Alloy$\,$1 (b) Alloy$\,$2 (c) Alloy$\,$3 (d) CM247LC when strained at 10$^{-2}$$\,$s$^{-1}$ and summary of their (e) flow stress vs temperature and (f) engineering strain to failure vs temperature.}
	\label{figure5}
\end{figure}

Figure~\ref{figure5} contains the data showing the variation of flow stress and ductility as a function of temperature for the 4 compositions following HT$\,$3. Consistent with the observed Nb and Ta-rich $\gamma^\prime$ observed by APT, Alloy$\,$3 exhibited the greatest flow stress of the 4 compositions up to 900$\,$$^\circ$C. Across the three novel alloy compositions, the flow stress increased with greater (Nb+Ta)/Al ratio. This occurred because Nb and Ta increases the energies of planar faults such as the anti-phase boundary (APB) and intrinsic/extrinsic stacking faults, which in turn inhibited $\gamma^\prime$ shearing \cite{crudden2014modelling}. Alloys$\,$1 and 2 -- with reduced Nb and Ta content -- displayed a corresponding reduction in strength across the temperature range tested. The increased strength of Alloy$\,$3 came with a corresponding reduction in ductility in the 700$\,$$^\circ$C-1100$\,$$^\circ$C temperature regime. A ductility dip was observed in the 4 compositions at 1000$\,$$^\circ$C, which is also observed down to 800$\,$$^\circ$C in Alloy$\,$3 and CM247LC. At the higher temperatures of 1000$\,$$^\circ$C and 1100$\,$$^\circ$C, CM247LC exhibited greater flow stress by $\sim$50$\,$MPa. These findings -- based on testing at a  rapid strain rate of 10$^{-2}$s$^{-1}$ -- do not account for the influence of any oxidation-assisted cracking \cite{nemeth2017environmentally}. Further investigation of the alloy properties taking into account the influence of oxidation-assisted cracking is carried out in the discussion section.

\subsubsection{Creep resistance}
	
\begin{figure}[H]
	\centering
	\includegraphics[width=0.9\columnwidth]{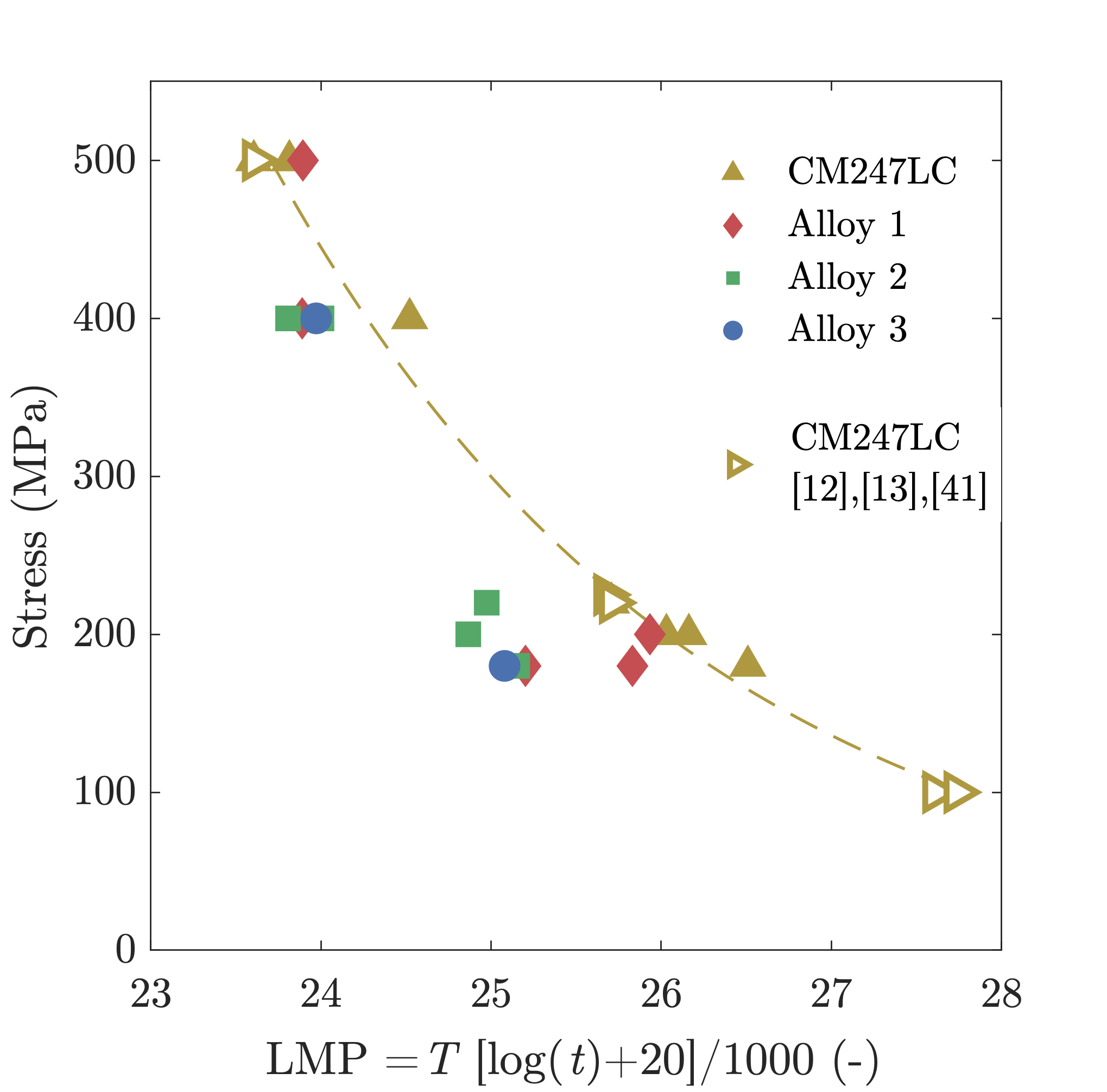}
	\caption{Plot of the stress level vs Larson-Miller parameter for the compositions creep tested with the build directon as the tensile axis with data from the literature \cite{zhou2020development, tang2021alloys, carter2013selective} superimposed on the results of this study.}
	\label{figure6}
\end{figure}

Figure~\ref{figure6} shows that the creep performance of CM247LC following HT$\,$3 exceeded that of the novel alloy variants across the range of temperatures and stresses tested by plotting the stress level vs the Larson-Miller parameter (LMP). The factors conferring CM247LC with increased creep life relative to the novel alloy variants are likely (i) the greater $\gamma^\prime$ volume fraction \cite{murakumo2004creep, zhu2012model} as well as (ii) the substantially higher C and B contents \cite{kontis2016effect, kontis2017role, despres2021role}. Good agreement was observed between the creep performance of L-PBF produced CM247LC -- tested parallel to the build direction -- reported here and in the literature \cite{zhou2020development, tang2021alloys, carter2013selective}. The superiority of CM247LC was reduced under conditions of higher stress and lower temperatures where APB shearing is the dominant deformation mechanism, as opposed to the low stress and high temperature conditions where dislocation climb bypass dominates \cite{smith2016creep, barba2017microtwinning}.
	
The creep resistance of CM247LC was greater than that of the novel alloys despite the extensive cracking incurred during processing. This performance was maintained as a result of the directionality of the cracks, which were planar in form and parallel to the build direction as shown in Figure~\ref{figure2}. This meant the load state on the process induced cracks during the testing performed in this study was a combination of Mode II and Mode III, in-plane shear and out-of-plane shear respectively. If the tensile axis of the mechanical testing performed had instead been perpendicular to the build direction, the load state on the internal cracks of CM247LC would be Mode I -- tensile stress normal to the crack plane -- which would likely increase the detrimental effect of the processing induced defects \cite{han2020effect}.

\newpage	
	
\subsection{Oxidation behaviour}
	
\begin{figure}[H]
	\centering
	\includegraphics[width=1\columnwidth]{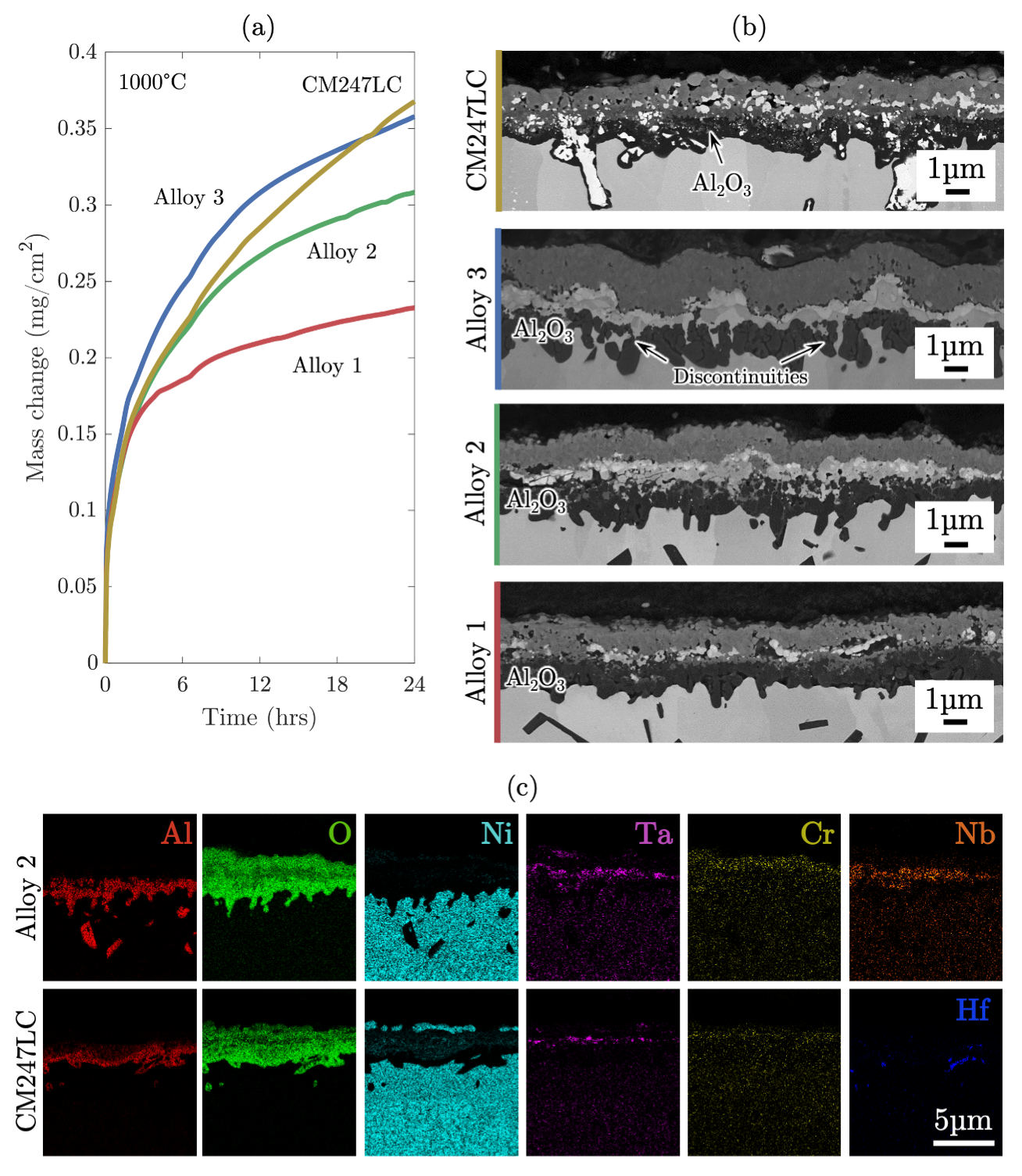}
	\caption{(a) Mass change vs time during isothermal thermo-gravimetric testing at 1000$\,$$^\circ$C. (b) BSE micrographs of the oxide developed on each composition following testing. (c) EDX maps showing the distributions of elements in the oxide layers formed on Alloy$\,$2 and CM247LC.}
	\label{figure7}
\end{figure}

A combination of TGA and post test characterization shown in Figure~\ref{figure7} demonstrates that the (Nb+Ta)/Al ratio had a substantial influence on the oxidation behaviour. Alloy$\,$1 had the greatest oxidation resistance on account of the stable continuous alumina layer formed. The total mass gain of the Alloy$\,$1 specimen after 24$\,$h was $\sim$$\,$60$\,$\% less than that of CM247LC and Alloy$\,$3. This is consistent with its having the highest Al content of the 3 novel alloy variants and thus most preferential thermodynamics and kinetics for the formation of a stable and continuous Al$_2$O$_3$ layer. This is evidenced by TGA and in the BSE micrographs  which show increasing mass gain and increasing discontinuity of the Al$_2$O$_3$ layers with decreasing Al content. In Alloy$\,$3 discontinuities are observed every $\sim$5$\upmu$m.
	
Characterization by EDX shed light on the nature of the scale formed in the novel alloy variants. The new compositions exhibited a thin outer Ni, Co, O rich oxide layer hypothesized to be (Ni$_{1-x}$Co$_x$)O, with a thicker Cr and O-rich Cr$_2$O$_3$ layer beneath, followed by a thin Nb, Ta-rich layer,  Al$_2$O$_3$, and AlN precipitates in the Al-depleted zone. While the overall combined oxide layer thicknesses of the 3 novel compositions were not statistically distinguishable, the Nb, Ta-rich layer was qualitatively observed to be thinner in Alloy$\,$1 -- the low Nb+Ta variant. AlN precipitated below sub-scale in Alloys$\,$1-3 was due to N uptake during testing. After 24$\,$h Alloys$\,$1-3 all exhibited relatively continuous Al$_2$O$_3$ layers, suggesting that the N uptake of the material occurred early on before the Al$_2$O$_3$ layers was established.

Discontinuous NiO oxide formed on the outer surface of CM247LC, followed by spinel, Cr$_2$O$_3$,  Ti/W-rich oxides, and lastly HfO$_2$ which was found throughout the oxide layers, having large concentration in the innermost Al$_2$O$_3$ oxide layer in particular. The average sizes of the Al depletion zone in Alloy$\,$1 and CM247LC (2.9$\,$$\upmu$m and 3.3$\,$$\upmu$m) were approximately half those of Alloy$\,$3 and Alloy$\,$2 (5.9$\,$$\upmu$m and 6.2$\,$$\upmu$m),  the standard deviations ranged from 0.4$\,$$\upmu$m to 0.6$\,$$\upmu$m. Despite its greater Al content, CM247LC is less resistant to oxidation than Alloys$\,$1-3 within this relatively short test period. This is likely due to its greater Ti content, which has been shown to have a deleterious effect on the oxidation resistance \cite{nemeth2018influence, kim2021regression}.
	
\section{Discussion}
	
The above results indicate that the new alloy and its variants show substantially improved processability and comparable performance in comparison with CM247LC.  However, key elements of these results warrant further analysis and are considered in what follows, namely, (i) the mechanism of super-solvus embrittlement, (ii) the resistance to oxidation-assisted cracking which is assuming greater and greater importance in industrial requirements, and (iii) the implications of the influence of the (Nb+Ta)/Al ratio on alloy design.
	
\subsection{On the mechanism of super-solvus ductility loss}
	
\begin{figure}[H]
	\centering
	\includegraphics[width=1\columnwidth]{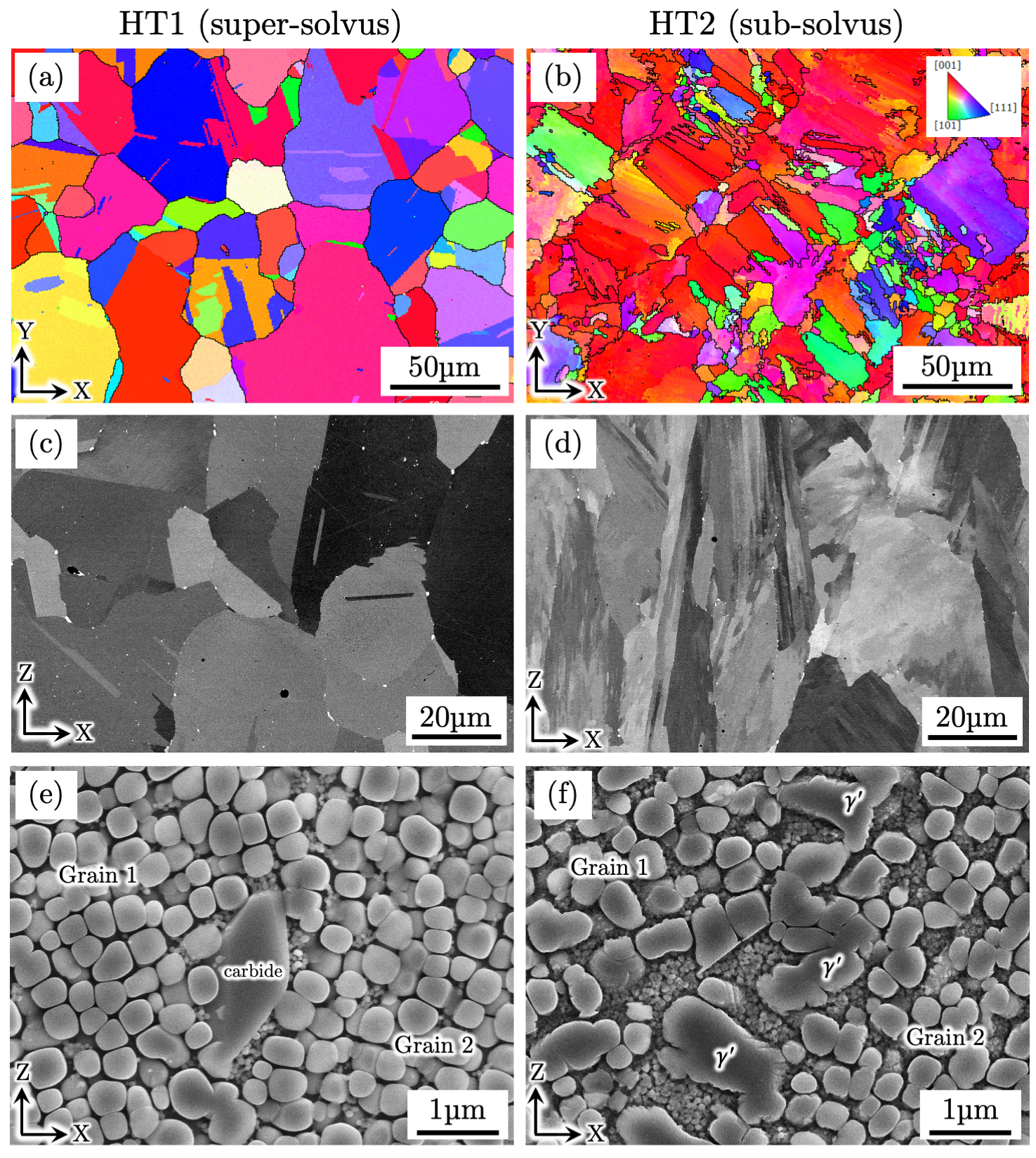}
	\caption{Microstructural analysis illustrating the cause of the super-solvus embrittlement of Alloy$\,$2. For super-solvus and sub-solvus respectively showing (a-b) EBSD inverse pole figure along the z-axis on the XY plane (c-d) BSE micrographs on the XZ plane (e-f) SE micrographs of the grain boundary microstructures on the XZ plane (where $\gamma$ matrix has been etched away).}
	\label{figure8}
\end{figure}

Characterization of the microstructure sheds light on the mechanism by which the embrittlement of Alloy$\,$2 following super-solvus heat treatment took place. The microstructures developed differ in terms of grain structure, carbide distributions, and local grain boundary microstructure. When heat treated above the $\gamma^\prime$ solvus, $\gamma^\prime$ is fully dissolved and recrystallization occurs; the growth of large equiaxed grains -- exemplified by EBSD and BSE in Figure~\ref{figure8} -- gives rise to long straight grain boundaries perpendicular to the tensile axis. In contrast, the sub-solvus heat treated microstructure remains serrated with the small grains consistent with the as-printed microstructure. Carbide coarsening occurred during both heat treatments, but takes place to a greater extent under super-solvus conditions after which large $\sim$1$\upmu$m diameter carbides were observed only on grain boundaries and often at grain boundary triple points. These grain boundary carbides have been shown to have a deleterious effect on ductility due to their pinning effect and ultimately decohesion of the carbide-matrix interface \cite{he2005effect, yang2011high}. After sub-solvus heat treatment the grain boundaries were decorated with some large and many fine $\gamma^\prime$ precipitates. This arises since $\gamma^\prime$ formers segregate to grain and cell boundaries during the printing process \cite{tang2020effect}, and promote the initial $\gamma^\prime$ precipitation when heated to 1080$\,$$^\circ$C which then coarsens during the isothermal holding, during cooling the fine $\gamma^\prime$ nucleate. The precipitation of fine $\gamma^\prime$ at the grain boundary during sub-solvus heat treatment of and its subsequent preservation of ductility has been previously reported \cite{joseph2017influence}. The distribution of $\gamma^\prime$ in the grain interior following super and sub-solvus heat treatments are quantitatively indistinguishable. As such, it is concluded that (i) the texture and (ii) local grain boundary microstructure are the key factors causing the embrittlement incurred by Alloy$\,$2 following super-solvus heat treatment.

\subsection{On the resistance to oxidation-assisted cracking (OAC)}
	
\begin{figure}[H]
	\centering
	\includegraphics[width=0.9\columnwidth]{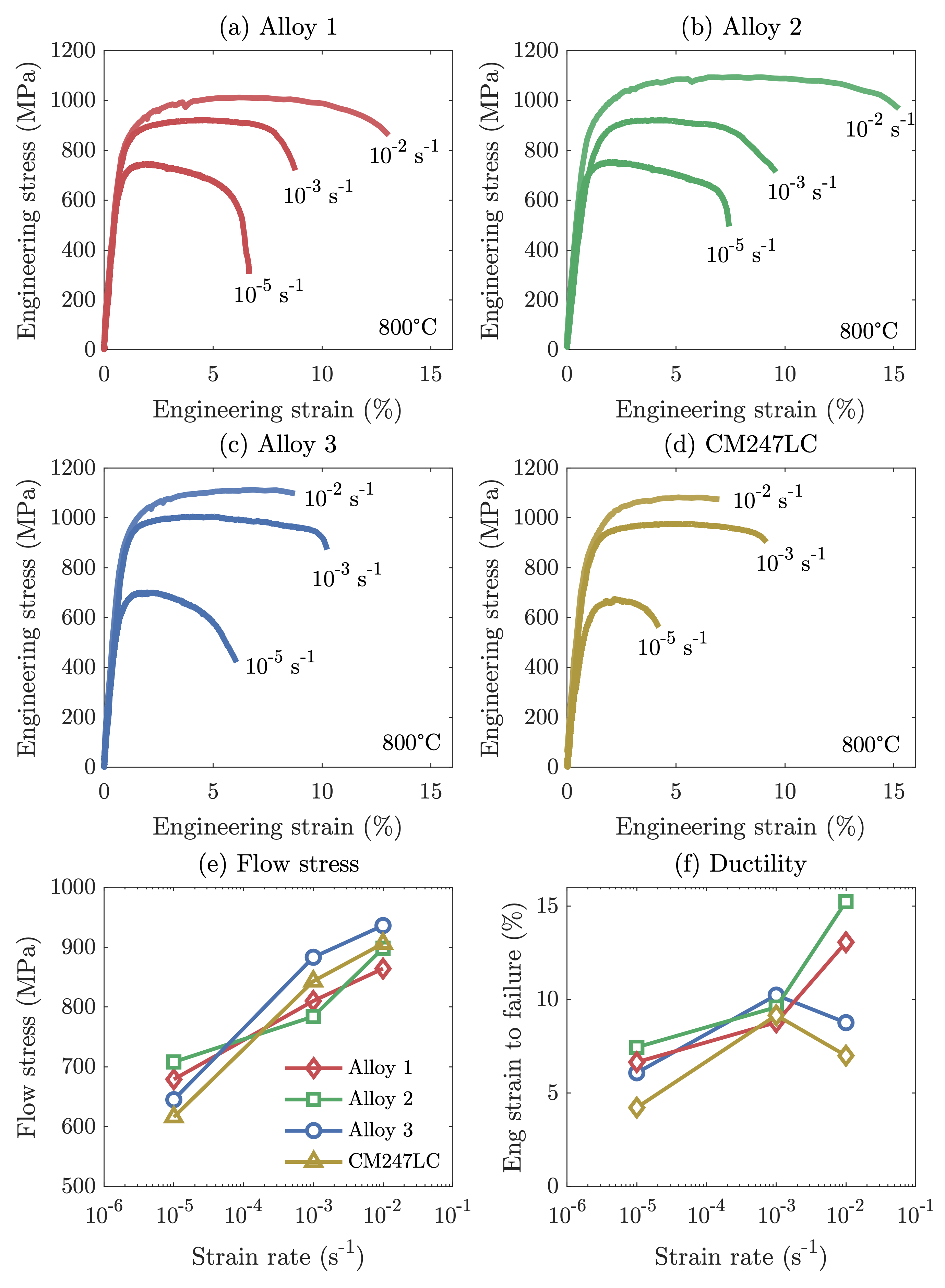}
	\caption{Strain rate dependence of the tensile response after HT$\,$3 for (a) Alloy$\,$1 (b) Alloy$\,$2 (c) Alloy$\,$3 and (d) CM247LC illustrating the their susceptibility to oxidation-assisted cracking at 800$\,$$^\circ$C. Summary of the strain rate sensitivity of (e) the flow stress and (f) the ductility.}
	\label{figure9}
\end{figure}

Alloy$\,$3 exhibits the highest strength and Alloy$\,$1 demonstrates the superior oxidation resistance, but the question is now asked; which is more desirable given the effect of oxidation on tensile properties? After all, it has been shown in the literature that superalloys strained at slow rates for longer periods of time exhibit oxidation at crack tips \cite{nemeth2017environmentally}; this accelerates crack propagation and substantially deteriorates the material properties and is of particular interest under dwell fatigue crack propagation conditions \cite{viskari2013intergranular}. When deformed at a rate of 10$^{-5}$$\,$s$^{-1}$ the alloys fracture after $\sim$2$\,$h. Each alloy exhibits a loss of approximately half their strength and ductility when deformed at this slower rate, see Figure~\ref{figure9}. This is consistent with the effects of oxidation as reported in the literature \cite{evans2009mechanism, lafata2018oxidation}.
	
Alloy$\,$2 displayed the most resistance to oxidation-assisted cracking, having the greatest flow stress $\sim$705$\,$MPa and ductility $\sim$8$\,$\% under uniaxial tension at a strain rate of 10$^{-5}$$\,$s$^{-1}$. Neither the most oxidation resistant high Al variant or the strongest -- high Nb+Ta variant -- performed as well as Alloy$\,$2 under these conditions. Thus, it appears the moderate ratio (Nb+Ta)/Al was most effective, with sufficient Nb and Ta to increase the $\gamma^\prime$ APB energy for strength as well as the Al to have oxidation resistance. The measured flow stress of CM247LC under a strain rate of 10$^{-2}$$\,$s$^{-1}$ was $\sim$900$\,$MPa, which dropped to $\sim$610$\,$MPa at 10$^{-5}$$\,$s$^{-1}$ given the contribution of oxidation. It is noted that the processing induced defects in CM247LC may have contributed to the decrease in properties at slower strain rate.
	
These findings highlight that despite initial tensile tests showing that the strategy of designing stronger alloys by substituting Al in $\gamma^\prime$ for Nb and Ta, a more holistic perspective indicates that maintaining sufficient Al content is vital. Overall, a balance of (Nb+Ta)/Al ratio must be struck to mitigate oxidation-assisted cracking. 

\newpage
	
\subsection{Composition-processing-performance trade-offs}
	
\begin{figure}[H]
	\centering
	\includegraphics[width=0.7\columnwidth]{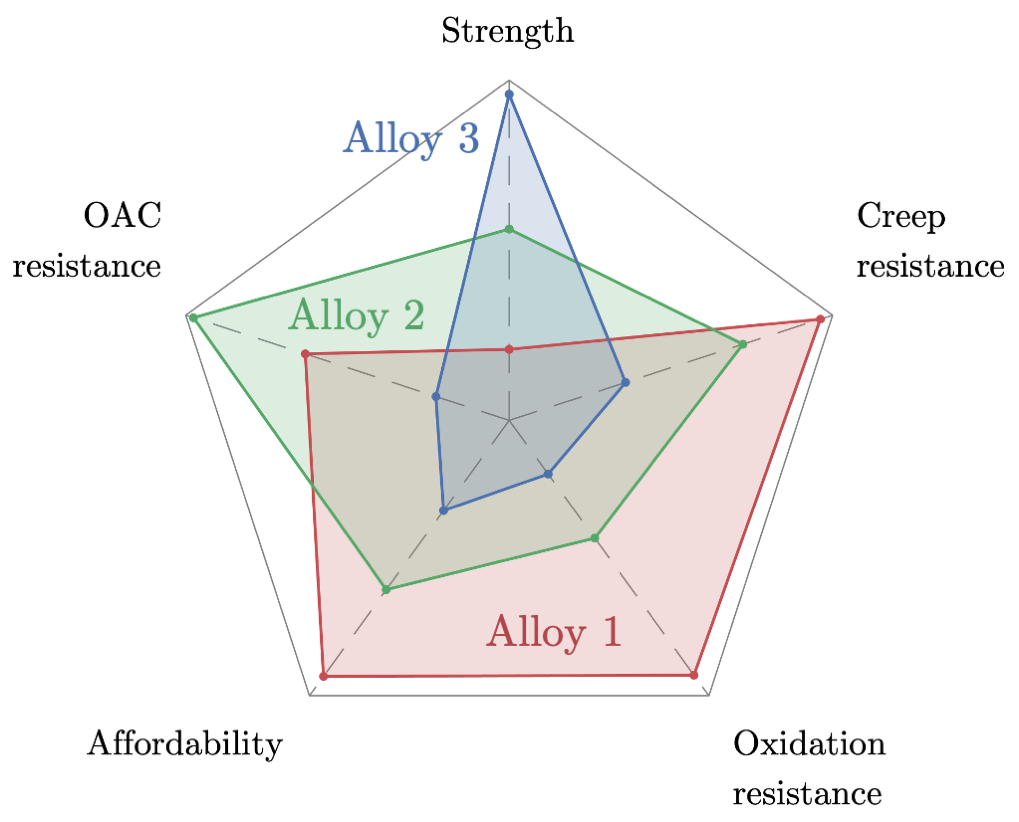}
	\caption{Spider plots summarizing the alloy design trade-offs with varying (Nb+Ta)/Al ratio.}
	\label{figure10}
\end{figure}

Here, five metrics are considered to summarize the performance trade-offs when the (Nb+Ta)/Al ratio is varied: strength, creep resistance, oxidation resistance, OAC resistance, and affordability. Maximizing each metric is ideal, the metrics are defined as follows; strength: the flow stress at 800$\,$$^\circ$C (MPa), affordability: the reciprocal of cost (kg/\$), creep resistance: the LMP at 180$\,$MPa, oxidation resistance: reciprocal of mass gain during isothermal oxidation, OAC resistance: the flow stress at 800$\,$$^\circ$C when strained at 10$^{-5}$$\,$s$^{-1}$. The metric values for the 3 alloy variants are summarized in Table~\ref{table3}. Figure~\ref{figure10} summarizes the trade-offs of these metrics in spider plots, the scaling of relative metrics is linear. Contrasting Alloy$\,$1 and Alloy$\,$3 shows that achieving the same $\gamma^\prime$ fraction through additions of either Al or Nb+Ta yields oxidation resistance or strength respectively. Given the Al is more readily available than Nb or Ta, the strength of Alloy$\,$3 also comes with greater cost. Alloy$\,$2 maintains a balance of strength and oxidation resistance and in doing so achieves good resistance to oxidation-assisted cracking relative to Alloy$\,$1 and Alloy$\,$3.

Direct comparison of CM247LC with the new compositions must be done, taking into account the application. The superior creep performance of CM247LC, which is attributed to its more elevated solid solution and interstitial content inhibiting the mobility of dislocations through the $\gamma$ channels comes at the price of processability, as the C and B content makes it susceptible to cracking \cite{despres2021role}. While it does retain superior creep properties despite the cracking it incurred during processing, these cracks may make it unsuitable on account of their deleterious effects for fatigue crack growth rate. In the mid-termperature regime -- where the shearing of $\gamma^\prime$ is the dominant deformation mechanism -- the novel compositions marginally exceed the performance of CM247LC. However, it is noted the heat treatment conditions applied in this work were selected on the basis of the $\gamma^\prime$ solvus temperatures of the new alloy variants and not CM247LC, and that there may be scope to optimize the heat treatment of CM247LC processed by AM \cite{munoz2016effect}.

The strength and oxidation resistance as well as the processability and creep trade-offs highlighted here offer valuable insights into alloy design. Esoteric amplification of any given material property at the expense of others may have unforeseen ramifications. Hence,  design must be undertaken with all the failure and damage mechanisms in mind, going past isothermal creep testing to assess alloy performance under conditions such as thermal-mechanical fatigue and dwell fatigue where oxidation-assisted cracking plays a role.

\begin{table}[H]
	\centering
	\caption{Spider plot performance values.}
	\begin{tabular}{@{}cccccc@{}}
		\toprule
		Alloy   & \begin{tabular}[c]{@{}c@{}}Strength  \\  (MPa)\end{tabular}  & \begin{tabular}[c]{@{}c@{}}Creep  \\ Resistance \\ (LMP)\end{tabular} & \begin{tabular}[c]{@{}c@{}}Oxidation \\ Resistance \\(24$\,$$\mathrm{h}/\mathrm{cm}^{2}/\mathrm{mg}$) \end{tabular} & \begin{tabular}[c]{@{}c@{}} Affordability \\ (kg$/\$$)  \end{tabular} & 	\begin{tabular}[c]{@{}c@{}}OAC \\ Resistance (MPa)  \end{tabular}
		
		\\ \midrule
		Alloy$\,$1 & 864           & 25.20                                                               & 4.3                                                             & 3.37              & 679                                                                                                      \\
		Alloy$\,$2 & 898           & 25.15                                                               & 3.3                                                             & 3.17             & 708                                                                                                   \\
		Alloy$\,$3 & 936           & 25.08                                                               & 2.9                                                             & 3.00              & 645                                                                                               \\\bottomrule
	\end{tabular}
	\label{table3}
\end{table}

\newpage

\section{Summary and Conclusions}
	
In this body of work, the processability and performance of three variants of a novel Ni-based superalloy have been assessed and contrasted with the heritage alloy CM247LC. The following specific conclusions have been drawn:
	
\begin{enumerate}
		
		\item The new grade of superalloy in its three (Nb+Ta)/Al variants is shown to be processable; processing-related cracking has not been detected by either optical microscopy coupled with stereology or else high resolution x-ray computed tomography. Conversely, the benchmark CM247LC alloy exhibited extensive cracking following L-PBF.
		
		\item The response of the as-printed material to heat treatment is critically important. Sub-solvus heat treatment is effective for the maintenance of ductility and strength, due to a fine decoration of $\gamma^\prime$ at the grain boundaries. Heat treatment conditions absent solutionizing need to be chosen carefully -- super-solvus heat treatment causes embrittlement due to the texture of the recrystallized grain structure and the local microstructure of the grain boundaries which then consist of $\gamma$ and coarse blocky carbides. 
		
		\item Increasing the (Nb+Ta)/Al ratio at a fixed $\gamma^\prime$ fraction increases the yield stress of the alloy by increasing the Nb and Ta content in $\gamma^\prime$, consistent with enhanced energies of planar faults. However, this comes at the expense of the alloy cost, oxidation -- and lastly -- the oxidation-assisted cracking resistance -- which is of increasing importance for the intended applications. It is highlighted that this ratio can be tailored to match the application of the alloy.
		
		\item This work serves as guidance for future alloy design and demonstrates that it is possible to identify compositions of superior yield stress, oxidation resistance, and oxidation-assisted cracking resistance which are nonetheless processable.
		
\end{enumerate}

\newpage	

\section{Acknowledgments}
	
The financial support of this work by Alloyed Ltd. as well as the the The Natural Sciences and Engineering Research Council of Canada (NSERC) in the Chemical, Biomedical and Materials Science Engineering division award number 532410. The authors acknowledge funding from Innovate UK, under project number 104047, specifically the Materials and Manufacturing Division, as well as funding for the National X-ray Computed Tomography (NXCT) grant code EP/T02593X/1 from the Engineering and Physical Sciences Research Council (EPSRC). The authors acknowledge Prof. Ian Sinclair, Dr. David Crudden, and Dr. Andr\'{e} N\'{e}meth for their support and advisory roles regarding this body of work.
 
\section{References}
	
	\bibliographystyle{spbasic-unsort}
	\bibliography{GhoussoubAM}

\end{document}